\documentclass[AMA,LATO1COL]{WileyNJD-v2}

\usepackage{algorithm,algpseudocode}
\usepackage{colortbl}
\usepackage{comment}

\articletype{Methodological Article}

\received{XX May 2020}
\revised{}
\accepted{}

\begin{document}

\title{Evolving Reference Architecture Description: \\Guidelines based on ISO/IEC/IEEE 42010 
}

\author[1,3]{Edilson S. Palma}

\author[2]{Elisa Yumi Nakagawa}

\author[1]{Débora Maria B. Paiva}

\author[1]{Maria Istela Cagnin}

\authormark{Edilson S. Palma \textsc{et al}}

\address[1]{\orgdiv{UFMS}, \orgname{Federal University of Mato Grosso do Sul}, \orgaddress{\state{Campo Grande}, \country{Brazil}}}

\address[2]{\orgdiv{USP}, \orgname{University of São Paulo}, \orgaddress{\state{São Paulo}, \country{Brazil}}}

\address[3]{\orgdiv{IFMS}, \orgname{Federal Institute of Mato Grosso do Sul}, \orgaddress{\state{Coxim}, \country{Brazil}}}

\corres{
Maria Istela Cagnin, College of Computing, Federal University of Mato Grosso do Sul, Zip Code 79070-900, Campo Grande, Brazil\\
\email{istela@facom.ufms.com} \\ 
}


\abstract[ABSTRACT]{
The architectural design of software systems is not a trivial task, requiring sometimes large experience and knowledge accumulated for years. Reference architectures have been increasingly adopted as a means to support such task, also contributing to the standardization and evolution of these systems. 
Although considerable time and effort are devoted to design these architectures, an outdated description is still found in several of them and, as a consequence, resulting in their non-continuation. 
This article presents guidelines to evolve the description of reference architectures, considering different types of stakeholders and required tasks.
To complement our statement that the guidelines are correct by construction as they were grounded in widely known international standard ISO/IEC/IEEE 42010 and literature, we also briefly present a qualitative analysis comparing the guidelines with an ad hoc way (commonly occurred in reference architectures).
We believe solutions like these guidelines are necessary and could further contribute to the sustainability and longevity of reference architectures.}



\maketitle

\section{Introduction}
\label{sec:intro}

As an earlier step of the software development process, the architectural design deals with the establishment of software architectures that describe software elements, externally visible properties of these elements, and relationships among them \cite{Bass:2012:SAP:2392670}. The establishment of a suitable architecture that meets functional and non-functional requirements is not a trivial task, requiring for that technical knowledge, experience, and skill accumulated for years.
In this scenario, reference architectures have been increasingly adopted as a means to facilitate the architectural design of software systems, promoting the reuse of architectural decisions, good practices, software elements, and so on, and generally speaking, the knowledge accumulated about how to architect systems in a given domain \cite{Nakagawa2014} \cite{Zhao2003}. Good examples of these architectures are Autosar\footnote{\url{https://www.autosar.org}} (for automotive sector), ARC-IT\footnote{\url{https://local.iteris.com/arc-it/index.html}} (for cooperative and intelligent transportation), and RAMI 4.0 \footnote{\url{https://www.plattform-i40.de/PI40/Redaktion/EN/Downloads/Publikation/rami40-an-introduction.html}} (for Industry 4.0).
In particular, the main difference between software architecture and reference architecture is that software architecture is a design solution for a specific software system; on the other hand, reference architecture offers a high-level design solution for a class of similar software systems belonging to a given domain. Due to this, reference architecture is more abstract than software architecture and must be instantiated and configured to attend the specificities of software being built (ISO/IEC/IEEE 42010 \cite{iso42010}; Nakagawa et al. \cite{Nakagawa:2011:RAP:2041790.2041818}; Galster and Avgeriou \cite{galster2011empirically}; Bass et al. \cite{Bass:2012:SAP:2392670}; Nakagawa et al. \cite{Nakagawa2014}).

A wider adoption of reference architectures depends directly on how appropriate their description\footnote{In this work, the terms architectural description, architectural documentation, and architectural representation have the same meaning.} is. Initiatives already exist to build adequate descriptions \cite{Guessi2011}; however, a number of architectures still have an outdated description in the sense that they were proposed, sometimes presented through some publications and, after that, such description is not updated, even those ones where their related application domain continually evolve. Evolution of reference architectures is necessary to comply with external agents (such as new laws, emergent technologies, new target platforms, new architectural styles and patterns) or internal issues (such as change of the business rules, new non-functional requirements, improvements/refinements in requirements), resulting in non-sustainable reference architectures \cite{Venters18SSRP}. For instance, Autosar does not still support the development of automotive systems for 100\% autonomous cars. Thus, due to all the demands of the market/business, there is a need to soon evolve Autosar.
Moreover, despite the importance of a systematic way to evolve the description of reference architectures, to the best our knowledge, none specific work has been conducted in that direction. 

The main goal of this article is to present guidelines to systematically evolve the description of reference architectures.
To establish these guidelines, problems and shortcomings found in reference architecture descriptions were identified based on a checklist of reference architectures assessment, called FERA (Framework for Evaluation of Reference Architectures) \cite{santos2013checklist} and, thereafter, were analyzed and summarized. Following, based on ISO/IEC/IEEE 42010 \cite{iso42010} (an international standard for software architecture description) and on works about software architectures evolution \cite{Sadou2005} \cite{Wang20132701} \cite{Ding2001191} adapted to the reference architectures context, we established our guidelines, which 
guide ``what to do'', ``how to make'', ``how to represent'', and ``which evolution tasks and rules'' are needed.
Due to this large foundation, we believe the guidelines are ``correct by construction''\footnote{``Correct by construction'' is commonly used in formal verification techniques to demonstrate that a system design is correct with respect to its specification. We ``lent'' this term to say our guidelines are correct with respect to what the software architecture community has required to perform architectural descriptions.}, but we additionally present a brief qualitative analysis to show the ease and viability to use these guidelines. Results pointed out they can more efficiently guide and facilitate the evolution of reference architecture description in comparison with 
an ad hoc way, i.e., the way how descriptions have been commonly updated. 
We believe these guidelines could be considered as a first and motivating step to systematize the evolution tasks and, as a consequence, they could contribute to become reference architectures sustainable along the time.

The remainder of this article is organized as follows. Section \ref{sec:back} presents a background related to our work and also discusses the related work concerned with the evolution of software architectures; some of them adapted in this work to the context of reference architectures. Section \ref{sec:diretrizesdeevolucao} presents our guidelines and the research method to establish them. Section \ref{sec:estudoempirico} briefly presents both the planning and conduction of the qualitative analysis. Section \ref{sec:discussion} discusses on our findings and limitations, and presents perspectives for future works.
Finally, Section \ref{sec:conclusion}  concludes this work.

\section{Background and Related Work}
\label{sec:back}


Reference architectures contain knowledge about how to design concrete architectures for a given application domain. They include business rules, architectural styles and patterns that address quality attributes, best practices for software development (e.g., architectural decisions, domain constraints, legislations and standards), and software elements that support the development of software systems for that domain \cite{Nakagawa:2011:RAP:2041790.2041818}. In addition, they can increase the effectiveness of the concrete architectures of a domain, helping to avoid rework with problems already solved \cite{Cloutier:SYS20129}.


As a software artifact, reference architectures also require a systematized process to be designed. In the literature, there are some processes with this aim, such as ProSA-RA \cite{Nakagawa2014}, which have some steps in common: i) ``Architectural Analysis’’ that deals with the collection and analysis of information sources to understand the target domain and to establish the architectural requirements of the reference architecture; ii) ``Architectural Synthesis’’ that deals with the reference architecture design based on previously defined architectural requirements; and iii) ``Architectural Evaluation'' that assesses the quality of the reference architecture.

During the architectural synthesis of a reference architecture, its description is built based on: (i) the architectural requirements identified for the reference architecture; (ii) the concepts of the target domain; and (iii) the architectural styles and patterns. For supporting this, authors \cite{Guessi2014162} \cite{Regli2014} establish architectural viewpoints and views to better represent reference architectures, taking into account their stakeholders and nature  \cite{Nakagawa2014}. It is important to emphasize that each view is the representation of the reference architecture as a whole, according to the needs of stakeholders.

In particular, ISO/IEC/IEEE 42010 \cite{iso42010}, which addresses the architectural description and is also basis of some works about reference architecture description, does not present how the description should be made, neither what architectural description languages or views should be used, but indicates which elements should be considered to represent a software architecture.


The description of reference architectures  also needs to evolve mainly due to deficiencies and defects identified after an evaluation, for example, using assessment checklists as FERA checklist \cite{santos2013checklist}\footnote{The current version of FERA containing 118 questions divided in four groups was reported in a journal paper under review in this moment and, therefore, not available completely in this work.}, or after evolution of the reference architectures themselves in which new requirements arisen or were not previously elicited. In case of reference architecture evolution, new elements can be incorporated, existing elements can be adapted, eliminated, and replaced to meet the new needs of the stakeholders.


We conducted a search to find works that propose or use approaches, processes, guidelines or techniques to the evolution of descriptions of reference architectures; however, no studies were found. Then, we conducted another search to find works addressing software architectures evolution that were also considered as information sources to establish our guidelines. It is worth detailing three main studies \cite{Sadou2005,Wang20132701,Ding2001191}.

The FOCUS approach \cite{Ding2001191} assists the recovery of architecture from source code of object-oriented systems as well as evolution of such architectures. This approach is composed of two stages: architectural recovery and architectural evolution. The first one aims to understand and document the architectures considering their logical and physical characteristics. The second stage, which is of our interest, evolves architectures based on an evolution plan. The evolution involves components added or modified and control flow among components updated. If necessary, new evolution iterations are performed at a lower level of granularity to evolve components increasingly specific.

Another work \cite{Wang20132701} investigates architectural evolution in the context of agile methods. The authors report that in agile methods the source code represents the software architecture itself and that architectural evolution is not something formal. In addition, iteration is the core of the process of architectural evolution, since new stories and those that need to be modified are included. Each iteration deals with the definition of the architecture, the construction of the architectural prototype, the implementation and the verification of the architecture and, finally, the release of the architecture that corresponds to the delivery of functional software. The authors explain the following operations that can occur in the software architecture evolution: add components, modify components, delete components, combine components, and split components. To identify the most appropriate operation, it is necessary to analyze the complexity between the relationships of the original components and the new components, as well as the connection and interaction between the static components.

Another work proposed an evolution model, called SAEV (Software Architecture EVolution model) \cite{Sadou2005}, which describes and controls the software architectures evolution at different levels of granularity as well as the evolution of software systems based on it. In particular, this model focuses on static (which occurs during software architecture specification) and dynamic (which occurs during software architecture execution) evolutions of software architectures. For this, SAEV considers during the architectural evolution all the architectural elements proposed by ADL (Architectural Description Languages), such as components, connectors, interfaces, and configurations. Additionally, this model defines the evolution operations (addition, removal, modification or replacement of architectural elements), evolution rules (that indicate the event that must occur to start the rule, the conditions that must be satisfied to perform the operation when the event occurs, as well as the actions to be performed), and evolution strategies (that represent the set of evolution rules describing all evolution operations applicable to each architectural element).

It should be noted that the aforementioned works \cite{Sadou2005, Wang20132701, Ding2001191} consider similar operations to the architectural evolution of software architectures; however, each of them uses specific mechanisms to identify the most appropriate operation to each case of evolution.

\section{Guidelines to the Evolution of Descriptions of Reference Architectures}
\label{sec:diretrizesdeevolucao}



The guidelines intend to help software architects to systematically evolve reference architecture descriptions to minimize their deficiencies.  Besides that, these guidelines could increase the quality of such descriptions, aligned to the need of less effort and time to evolve such descriptions. The subsections following describe: (i) the information source used to specify the guidelines; (ii) how the structure and content of the guidelines were defined; and (iii) a roadmap and an example of their use.

\subsection{Information Sources}
\label{diretrizesTrabalhosElaboracao}

We initially analyzed works that proposed and used methods to evaluate reference architectures themselves to obtain the evolution requirements that could be considered during the evolution of a reference architecture description and that should be treated by the guidelines. From this analysis, we identified the possible shortcomings of reference architectures that lead to the evolution of their description. 

Three main works \cite{gallagher2000using} \cite{graaf2005evaluating} \cite{angelov2008towards} presented adaptations of software architectures assessment methods to the evaluation of reference architectures, focusing on architectural elements related to the quality of these architectures.
In addition to these works, the FERA checklist \cite{santos2013checklist} also proposed to evaluate reference architectures under the aspects of quality attributes and aspects related to architectural representation, i.e., FERA is based on ISO/EIC 42010. Due to such characteristics and its use in the assessment of several reference architectures belonging to academia and industry \cite {osshiro2018cambuci} \cite{PORTOCARRERO20173} \cite{Portocarrero:2015} \cite{rodriguez2015reference} \cite{duarte2015contribution}, FERA was selected as an information source to establish our guidelines.
FERA consists of four parts (Part 1 - General Information, Part 2 - Construction and Content, Part 3 - Appropriate Architectural Description and Part 4 - General Analysis and Conclusion) and each part contains questions answered by different types of stakeholders of reference architectures, such as software architect, domain expert, manager, developer, software quality assurance specialist, tester, and systems analyst (i.e., those interested  in the conception, construction, validation, and use of reference architectures). The possibility of evaluations conducted from the perspective of several stakeholders was another reason to use this checklist in our work.
Each question of FERA provided us requirements to specify our guidelines and to establish what each guideline should address to deal with shortcomings and defects in the descriptions of reference architectures.


Studies about software architectures description and evolution (discussed briefly in Section \ref{sec:back}) were also considered to built the guidelines, since no studies were found related to the evolution of descriptions of reference architectures. 
%
%
Moreover, a systematic literature review about reference architecture representation, previously conducted in 2011 by Guessi et al. \cite{Guessi2011}, was updated for this work that identified other 21 studies \cite{AndhariniDwi2015284,Szwed2015222, Branscomb201379, Braun201148,DeLaCruz2011305, Filho2015, Gringinger2012, Guessi2014162, Guessi:2015:TFD:2755567.2755571, Guessi2014, Turek2015, Kruize201375,  Lopes2011441, Ma2013197, Mafazi2014, Nakagawa2012297, Nakagawa2014, Paschke2012, Pereira2014185, Regli2014, Stocker2015}. These studies were used in our work to identify how to perform the reference architecture description, what supported us to propose our guidelines. To complement, another study of Guessi \cite{Guessi2013mestrado}, which defines a method to describe  reference architecture for embedded systems, was also used to identify architecture views and viewpoints to describe reference architectures.

\subsection{Structure and Content of the Guidelines}
\label{diretrizesElaboracao}

\begin{figure}[h]
    \centering
	\tiny
	\centering
         \includegraphics[width=175mm]{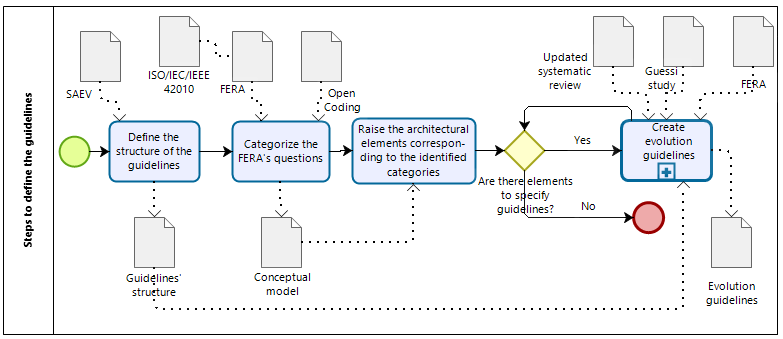}
	\caption{Steps performed to construct the guidelines for the evolution of reference architecture descriptions}
	\label{steps_guidelinesdefinition}
\end{figure}

Based on the information sources identified, the structure and content of the guidelines were defined, as described below and summarized in Figure \ref{steps_guidelinesdefinition}\footnote{Represented in Business Process Model and Notation (BPMN).}.

Firstly, the structure of the guidelines was defined. For this, we used as the basis SAEV \cite{Sadou2005}, including their important concepts: architectural element, evolution operation, evolution rule, and evolution strategy. These concepts were mapped in our work, respectively, as \textbf{element}, \textbf{evolution task}, \textbf{evolution rule}, and \textbf{evolution guideline}.

\textbf{Elements} represent the architectural elements that are considered for the evolution of reference architecture description (e.g., Variability element, ViewPoint Element, View Element, Instantiation Element, etc). Such elements are composed of one or more components (e.g., description of a viewpoint, representation block of a model belonging to a view, etc).

\textbf{Evolution tasks} are classified into three types: addition, removal, and modification; which respectively add, remove or modify elements and components in a reference architecture description. The definition of these tasks were based on works described in Section \ref{sec:back} \cite{Wang20132701, Ding2001191} and mainly the SAEV model \cite{Sadou2005}, which identifies \textbf{evolution rules} for each one of the evolution tasks.An evolution task is composed of one or more evolution rules, and such rules are represented based on the ECA (Event/Condition/Action) formalism, used in \cite{Sadou2005}. Each evolution rule has an \textit{event}, which specifies when the rule should be triggered; a \textit{condition}, which is evaluated when the event occurs, enabling the execution of the rule only if such conditions are true; and an \textit{action} that must be performed when the event occurs and the conditions are satisfied.

Moreover, an \textbf{evolution guideline} must provide a solution to solve any deficiency of a reference architecture description found by an assessment based on FERA (\textbf{``what to do''}). Knowing what needs to be done, it is essential to point out \textbf{``how to represent''} the solution in the reference architecture description, \textbf{``how to make''} to evolve a reference architecture description and which \textbf{artifacts} are important to be considered for the execution of each proposed guideline.

Hence, the structure to document each guideline is composed of an \textit{Identifier} (according to the format D\_{[}Element{]}\_{[} Sequential number{]}), the \textit{Element} (that represents the category of FERA's questions handled by the guideline), the \textit{Identifiers of FERA's questions} addressed by the guideline (according with the format {[}Group of FERA{]} - {[}Identifier of FERA's question{]}), \textit{What to do}, \textit{How to represent}, \textit{How to make}, the \textit{Evolution tasks and rules}, and the \textit{Artifacts involved} to carry out the evolution. In the case of the evolution rules, they present an \textit{Identifier} (in the format R-{[}Acronym of the guideline in lowercase{]}-{[}Sequential number{]}), the \textit{Event} that triggers the rule, the necessary \textit{Condition} to start the execution of the rule, and the \textit{Action} itself.

Afterwards,  we categorized the FERA’s questions. For this, all 118 questions of FERA were sorted systematically applying the Open Coding technique \cite{strauss1998basicsgrounded}. Each code was firstly identified by the first author of this paper based on concepts addressed for each FERA’s question. For instance, in the question \textit{``Is there traceability between the domain objectives and the technical solution?''} of FERA, the codes identified were DomainData and TechnicalSolution and in the question \textit{``Is there a clear description of which parts of the reference architecture are fixed and which parts are subject to instantiation in a concrete organization/context?''} the code was Variability. To accomplish this task, we used R software\footnote{https://www.R-project.org} and package RQDA\footnote{http://rqda.r-forge.r-project.org}, which helped us to label each question by inserting the codes of interest. Iteratively, each code generated was discussed, reviewed, and validated by all other authors. Consensus meetings were also held to discuss the codes extracted and their relationships.

It is noteworthy that it was necessary to create codes identified from the FERA’s questions that are beyond what is covered by ISO/IEC/IEEE 42010. These codes were concerned: (i) with important aspects not addressed by the standard (instantiation code), (ii) specific aspect of reference architectures (variability code),  and (iii) others elements and artifacts of reference architecture life cycle considered relevant by FERA to be evaluated (acquisition, test, quality, risk or threat, evolution, domain data, information source, module and technical solution codes).

\begin{figure*}[ht] \footnotesize
    \centering
	\tiny
	\centering
         \includegraphics[width=170mm]{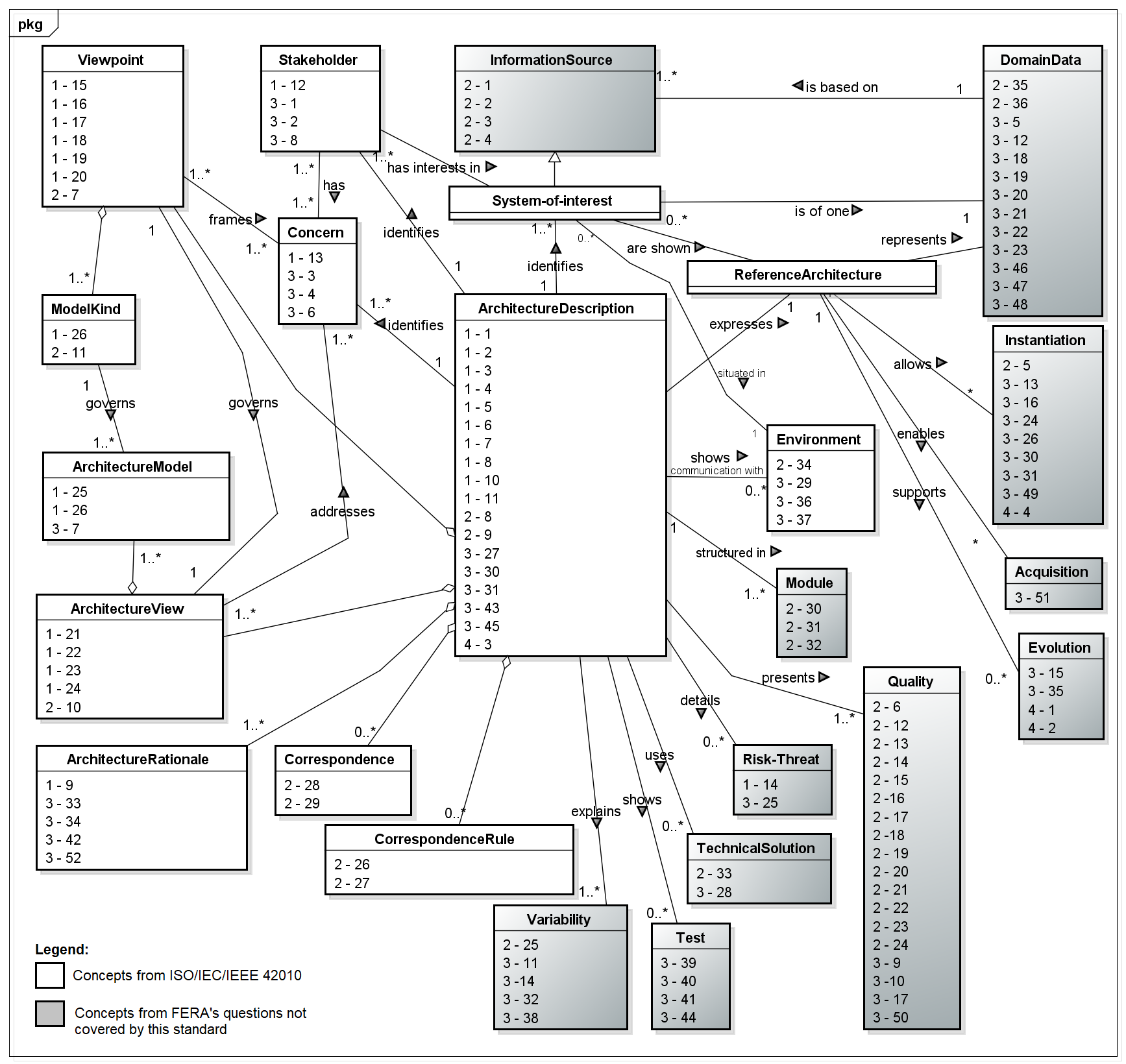}
	\caption{Conceptual model that presents the categories of FERA's questions}
	\label{modeloconceitual_iso42010_FERA}
\end{figure*}

Following, the conceptual model illustrated in Figure \ref{modeloconceitual_iso42010_FERA}\footnote{We represented the conceptual model using the notation of UML class diagram.} was elaborated containing 24 concepts. To build it, all identified categories were represented as concepts and shown in the diagram exhibited in this figure. While categories related to FERA’s questions and to ISO/IEC/IEEE 42010 are presented as classes in white background, other new categories are presented as classes in gray background. In addition, the identifier of the FERA’s questions (in the format ``Part-Question''), which were classified in each category, are presented in the class related to that category. For instance, in the category ``Viewpoint'', there are seven questions, including Question 15 of Part 1 (i.e., 1-15). This question is ``All selected viewpoints for the reference architecture were considered?'', which in fact is related to the category Viewpoint. Another example is the category DomainData that has 13 questions, where for instance Question 35 of Part 2 is ``Is the reference architecture in conformance with the requirements document?'', i.e., such question asks about domain data into the reference architecture.

Regarding the relationships among the concepts, they were established in accordance with the ISO/IEC/IEEE 42010. Relationships with the concepts obtained from FERA’s questions that are beyond what is covered by this standard were created based on the context of these questions. In particular, a relationship was included between two existing concepts, that is \textit{ArchitectureDescription} and \textit{Environment}, as there exist four FERA’s questions (whose identifiers are 2-34, 3-29, 3-36, 3-37) related to the reference architecture and its external environment and must be considered by its description.

The conceptual model was useful to understand the elements involved in the reference architecture description and that shall be carefully considered during its building and evolution. In addition, this conceptual model was very helpful during the guidelines construction, as it explains the elements that should be addressed by the guidelines and also contains the identifiers of the FERA’s questions. From these identifiers, all information about the questions was obtained to define the guidelines.

After categorizing the questions, we took the categories as base to raise the corresponding architectural \textbf{elements} to be evolved (according to the third activity of Figure \ref{steps_guidelinesdefinition}). Next, we started the elaboration of the guidelines for each raised architectural element, and detailed in Algorithm~\ref{guidelines-structure-and-content}. FERA’s questions corresponding to each element were analyzed to identify evolution requirements. Then, for each requirement, an \textbf{evolution guideline} was created.
For this, we find solutions (\textbf{what to do}) to solve them, i.e., what is necessary to be done in the description of reference architectures for that the corresponding questions are answered positively during a reference architecture evaluation. Questions associated with each architectural element predisposed to evolve and that share a similar solution were grouped into the same guideline. Hence, one or more guidelines may be associated with a single element.

\begin{algorithm}
\caption{Activity ``Create Evolution Guidelines'' steps}
\label{guidelines-structure-and-content}
\begin{algorithmic}[1]
\For{\textbf{each} architectural \textbf{``Element''} raised}
\State {Determine the evolution requirements based on checklist’s questions associated to category corresponding to element}
\For{\textbf{each} evolution requirement}
\State Create an evolution guideline and name it according to the format D\_{[}Element{]}\_{[} Sequential number{]}
\State Find solution \textbf{``What to do''} to solve the evolution requirement 
\State Determine \textbf{``How to represent''} the solution in the reference architecture description
\State Define \textbf{“How to make”} to evolve the reference architecture description
\begin{itemize}
\State Establish the \textbf{``Evolution tasks''} with the respective \textbf{``Evolution rules''}
\State Identify the \textbf{``Artifacts''} that should be considered to evolve the reference architecture description
\end{itemize}
\State Document the guideline based on structure defined in the first activity (Figure \ref{steps_guidelinesdefinition}) and on content defined in Steps 5 to 7.
\EndFor
\EndFor
\end{algorithmic}
\end{algorithm}

Subsequently, we analyzed \textbf{how to represent} the solution in the reference architecture description, based on the studies found in \cite{Guessi2013mestrado,AndhariniDwi2015284,Szwed2015222, Branscomb201379, Braun201148,DeLaCruz2011305, Filho2015, Gringinger2012, Guessi2014162, Guessi:2015:TFD:2755567.2755571, Guessi2014, Turek2015, Kruize201375,  Lopes2011441, Ma2013197, Mafazi2014, Nakagawa2012297, Nakagawa2014, Paschke2012, Pereira2014185, Regli2014, Stocker2015}, which contain works about views and viewpoints to represent reference architectures. Next, we defined \textbf{how to make} to evolve a reference architecture description. For this, we considered two steps (``Architectural Analysis'' and ``Architectural Synthesis'') commonly found in processes to build such architectures such as ProSA-RA, and that should be executed based on ``what to do'' and ``how to represent'' previously discussed. When it is necessary to obtain new information sources or to analyze existing ones, we indicate the execution of the ``Architectural Analysis'' step; otherwise, we recommend the execution of the ``Architectural Synthesis'' step. The first step contains activities that guide architects making it possible to examine the information source and the second step is executed with the support of \textbf{evolution tasks}, defined based on ``what to do'' and ``how to represent'', and provided by the guidelines to effectively carry out the evolution of the architectural description. In parallel, we identified important artifacts (such as architectural description and traceability matrix) that should be considered for the execution of each proposed guideline. Finally, the guideline is documented based on the previously defined structure.

Ultimately, we identified 52 guidelines associated to 22 elements\footnote{No guideline was associated with the categories \emph{System-of-interest} and \emph{Reference Architecture}, since they do not represent elements that make up the architectural description and, therefore, they are not elements of evolution interest.} together with their evolution rules, as illustrated in Table \ref{dVariability1} (that shows a given guideline) and Table \ref{rule_Variability} (that shows one of the evolution rules of this guideline). Other five guidelines are presented in Appendix A\footnote{\url{https://goo.gl/gqyF2o}}.

Regarding the guideline in Table \ref{dVariability1}, it deals with the \textbf{Variability} element in reference architecture description. This guideline points out that all elements of the reference architecture shall be mapped as mandatory or variable, aiming at expliciting and documenting the domain variability covered by reference architecture. For this, this guideline indicates several existing techniques that can be used by the architect during the description evolution, as well as how to make the evolution, the needed artifacts and the evolution tasks and rules that must be carried out. In a nutshell, \textbf{R-var-1} (presented in Table \ref{rule_Variability}) details that a Variability View must be created (whether it does not exist), a Model Kind must be chosen according to one of the adopted techniques to document variability, the Variability Model must be created according to Model Kind chosen, a suitable Viewpoint must be created (whether it does not exist) and the Variability View must be incorporated. Then, the Variability Model is added in the Variability View and the Traceability Matrix is updated. \textbf{R-var-2} indicates that the Variability Model and the Variability View must be removed and the Traceability Matrix must be updated. \textbf{R-var-3} specifies that the relevant modifications regarding variability must be incorporated in the Variability Model and the Traceability Matrix must be updated, if pertinent.

\begin{table*}[ht!]
\footnotesize 
	\centering
	\caption{Example of a guideline (Guideline D\_Variability\_1)}\
	\label{dVariability1}
	\begin{tabular}{|l|l|}
		\hline
		\rowcolor[HTML]{9B9B9B} 
		\textbf{Guideline}                                                                                     & \begin{tabular}[c]{@{}l@{}} D\_Variability\_1\end{tabular}                                                                                                             \\ \hline
		\rowcolor[HTML]{C0C0C0} 
		\textbf{Element}                                                                                     & Variability                                              \\ \hline
		\textbf{\begin{tabular}[c]{@{}r@{}}Applicable to \\FERA questions \end{tabular}}                  & \begin{tabular}[c]{p{0.8\textwidth}}2-25; 3-11; 3-14; 3-32; 3-38
        \end{tabular}                                                                \\ \hline
		\textbf{What to do}                                                                                  & \begin{tabular}[c]{p{0.8\textwidth}}
	            To present the mapping of reference architecture’s elements as mandatory or variable, respectively  representing the commonalities and variabilities of the domain covered by the reference architecture. For each variability, it is necessary to classify it. In addition, constraints should be defined between communalities and variability, if any.
		\end{tabular}        
		\\ \hline
		\textbf{\begin{tabular}[c]{@{}r@{}}How to represent\end{tabular}}  & 
		
		\begin{tabular}[c] {p{0.8\textwidth}}
		\textbf{Graphical}: Nakagawa et al. \cite{Nakagawa2014} propose a Variability View associated with a Crosscutting Viewpoint. Feature model \cite{kang1990} and orthogonal model \cite{pohl2005} are some of the existing techniques to represent the Variability View. These models allow to represent variabilities, communalities and constraints among the elements. Another way to represent variability, presented by Guessi et al. \cite{Guessi2014}, is through ADL (Architectural Description Language) SysML \cite{omgsysml}.
		\end{tabular}                                                                  \\\hline
		\textbf{How to make}                                                                                   & \begin{tabular}[c]{p{0.8\textwidth}}
		
		Execute step ``Architectural Synthesis'' based on evolution tasks of interest: Addition, Removal or Modification.
		\end{tabular} \\ \hline
	
		& \begin{tabular}[c]{p{0.8\textwidth}}
		Addition: R-var-1
\end{tabular}      \\ \cline{2-2} 
		
		& \begin{tabular}[c]{p{0.7\textwidth}}
		Removal: R-var-2
		\end{tabular}       \\ \cline{2-2} 
		\multirow{-3}{*}{\textbf{\begin{tabular}[c]{@{}r@{}} Evolution tasks \\
		and rules\end{tabular}}} & 
		\begin{tabular}[c]{p{0.8\textwidth}}
		Modification: R-var-3
		\end{tabular} \\ \hline
		\textbf{\begin{tabular}[c]{@{}r@{}}Artifact  involved \\  \end{tabular}}                              & \begin{tabular}[c]{p{0.8\textwidth}}Architectural Description and Traceability Matrix \end{tabular}  \\ \hline
	\end{tabular}
\end{table*}

\begin{table*}[ht!] 
\footnotesize{
	\centering
	\caption{Example of an evolution rule (One of the rules of guideline D\_Variability\_1)}
	\label{rule_Variability}
	\begin{tabular}{|l|l|}
		\hline
		\rowcolor[HTML]{C0C0C0} 
		\textbf{\begin{tabular}[c]{@{}r@{}}Evolution rule\end{tabular}} & \begin{tabular}[c]{p{0.8\textwidth}}
		R-var-1\end{tabular} \\ \hline
		\textbf{Event}
		          & Add Variability element. \\ \hline
		\textbf{Condition}                                                     & \begin{tabular}[c]{p{0.8\textwidth}} 
		--
		\end{tabular} \\ \hline

		\textbf{Action}                                                         & \begin{tabular}{p{0.8\textwidth}}
            - Create the Variability View.\\
            - Choose the Model Kind to represent the Variability View. \\
            - Create the Variability Model according to Model Kind chosen. \\
            - If there is no suitable Viewpoint to represent variability then \\
            \begin{tabular}{p{2.0\textwidth}}
                   - Create the Viewpoint. \\
            \end{tabular}
            - Else \\
            \begin{tabular}{p{2.0\textwidth}}
                   - Indicate the suitable existing Viewpoint.\\
            \end{tabular}
            - Incorporate the Variability View  into the Viewpoint.\\
            - Add the Variability Model into Variability View.\\
            - Update the Traceability Matrix.
\end{tabular}                                            \\ \hline
	\end{tabular}}
\end{table*}

\subsection{Context and Use of the Guidelines}
\label{diretrizesContextoUso}


Besides applying the guidelines during the evolution of reference architectures (in particular, on their description), they could be applied during the design of such architectures early providing direction that could reduce further time/effort for evolution.


To facilitate the use of the most appropriate guidelines, the index shown partially in Table \ref{exemplo-resumo-diretriz} (that shows two of 22 elements) was created and contains: (i) the element; (ii) the identifier of the guidelines; (iii) the FERA's questions that each guideline can be applied; and (iv) the evolution tasks and rules associated with each guideline. Besides that, the roadmap presented in Algorithm~\ref{roadmap} provides steps that guide the use of the guidelines. 


\begin{table*}[ht!] 
\footnotesize
\centering
\caption{Part of the index for facilitating the use of the guidelines}
\label{exemplo-resumo-diretriz}
\begin{tabular}{|c|c|c|c|l|}
\hline
\rowcolor[HTML]{C0C0C0} 
\textbf{Element}                                 & \textbf{Guideline Name}                                       & \textbf{\begin{tabular}[c]{@{}c@{}}Checklist's\\ questions\end{tabular}}              & \textbf{\begin{tabular}[c]{@{}c@{}}Evolution\\ 			Task\end{tabular}} & \multicolumn{1}{c|}{\cellcolor[HTML]{C0C0C0}\textbf{\begin{tabular}[c]{@{}c@{}}Evolution\\ 			Rule\end{tabular}}} \\ \hline
                                                 &                                                          &                                                                                   & Addition                                                             & R-fi-1; R-fi-2                                                                                                    \\ \cline{4-5} 
                                                 &                                                          &                                                                                   & Removal                                                              & -                                                                                                                 \\ \cline{4-5} 
\multirow{-3}{*}{Information Source} & \multirow{-3}{*}{D\_Information\_Source\_1} & \multirow{-3}{*}{\begin{tabular}[c]{@{}c@{}}2-1;\\ 			2-2; 2-3; 2-4\end{tabular}} & Modification                                                         & R-fi-3                                                                                                            \\ \hline
                                                 &                                                          &                                                                                   & Addition                                                             & -                                                                                                                 \\ \cline{4-5} 
                                                 &                                                          &                                                                                   & Removal                                                              & -                                                                                                                 \\ \cline{4-5} 
                                                 & \multirow{-3}{*}{D\_Instantiation\_1}           & \multirow{-3}{*}{2-5; 4-4}                                                        & Modification                                                         & \begin{tabular}[c]{@{}l@{}}R-vi-4; \\ R-ponvi-4\end{tabular}                                                      \\ \cline{2-5} 
                                                 &                                                          &                                                                                   & Addition                                                             & R-ins-1; R-ins-2; R-ins-4                                                                                         \\ \cline{4-5} 
                                                 &                                                          &                                                                                   & Removal                                                              & -                                                                                                                 \\ \cline{4-5} 
                                                 & \multirow{-3}{*}{D\_Instantiation\_2}           & \multirow{-3}{*}{3-13; 3-24}                                                      & Modification                                                         & R-ins-3; R-ins-5                                                                                                  \\ \cline{2-5} 
                                                 &                                                          &                                                                                   & Addition                                                             & R-ins-6                                                                                                           \\ \cline{4-5} 
                                                 &                                                          &                                                                                   & Removal                                                              & -                                                                                                                 \\ \cline{4-5} 
\multirow{-9}{*}{Instantiation}         & \multirow{-3}{*}{D\_Instantiation\_3}           & \multirow{-3}{*}{3-16}                                                            & Modification                                                         & R-ins-7                                                                                                           \\ \hline
\end{tabular}
\end{table*}

\begin{algorithm}
\caption{Roadmap to support the use of guidelines}
\label{roadmap}
\begin{algorithmic}[1]
\State{Find the element \emph{E}  addressed by the question that identified shortcomings in reference architecture during its evaluation with the FERA (for example, question \emph{Q}), observing the conceptual model of Figure~\ref{modeloconceitual_iso42010_FERA}.}
\State{Search for element \emph{E}  in the guidelines index.}
\State{Search for question \emph{Q} indicated in the element \emph{E}.}
\State{Select guideline \emph{D} that is applicable to the question \emph{Q}.}
\State{Select evolution tasks of guideline \emph{D} that will resolve the shortcoming identified by the question \emph{Q}.}
 \For{\textbf{each} evolution task 
 chosen}
\State Select the most appropriate evolution rules. At the moment, the comments associated with the answer of the question \emph{Q}, described by the evaluators, can be considered.
\EndFor
\end{algorithmic}
\end{algorithm}

\section{Qualitative Analysis}

\label{sec:estudoempirico}

The objective of the qualitative analysis was to compare our guidelines with an ad hoc approach to evolve the description of a service-oriented reference architecture for software assets repositories, the Cambuci \cite{osshiro2018cambuci}. For the participants of the group that used the ad hoc method, they received the description of Cambuci and the ISO/IEC/IEEE 42010. These participants could evolve the description of Cambuci using their knowledge of how to represent software architectures (obtained through the training provided and described in Section \ref{sec:planejamento} and also by consulting the ISO/IEC/IEEE 42010). The criteria used in the comparison were: (i) ease to use; (ii) intention to use; and (iii) efficiency with regard to completeness and correctness of the descriptions after evolved.


Firstly, an evaluation of Cambuci using FERA identified several deficiencies \cite {osshiro2018cambuci}. Considering three of them, three evolution requests were raised and emphasized during the execution of this qualitative analysis: 
\begin{itemize}
    \item R1: For each architectural view of Cambuci, identify possible stakeholders (Low complexity); 
    \item R2: Enable the architectural description of Cambuci to represent open decisions (Average complexity); and 
    \item R3: Modify architectural views of Cambuci to make them consistent (High complexity).
\end{itemize}

\subsection{Planning}
\label{sec:planejamento}

Eighteen graduate students in Computer Science of the Federal University of Mato Grosso do Sul, who were attending the course ``Software Development'' (Master's degree level) during the second semester of 2017 were invited to participate in this qualitative analysis. It is worth noting that six of them were practitioners and had been working in the industry for an average of eight years and performing the following roles: analyst, developer, and project manager. These students were distributed in two groups (\emph{G-Guidelines} and \emph{G-Ad-Hoc}). 
For this, we took into account the background of them, obtained through a profile form. We randomly and balancedly distributed the participants with knowledge and professional experience in software development in both groups aiming at obtaining balanced groups, resulting in two groups with nine participants.

One and a half hours training was firstly provided to both groups to equal the knowledge, what included concepts on software architecture and reference architecture, ISO/IEC/IEEE 42010, and a summary about representation of reference architectures. Specifically for \emph{G-Guidelines}, the following documents were provided: the description of Cambuci and guidelines; while for \emph{G-Ad-Hoc}, we provided
them the description of Cambuci and the ISO/IEC/IEEE 42010.
 
We also provided for both groups, general instructions as follows:

        \begin{itemize}
        \setlength\itemsep{-0.05em}
            \item \underline{Step 1}: Read and understand the Cambuci architectural description;
            \item \underline{Step 2}: Perform the three evolution requests:
            \begin{itemize}
            \setlength\itemsep{-0.4em}
                \item Describe the evolution tasks required to meet each evolution request; and
                \item Update the architectural description of Cambuci based on the described evolution tasks.
            \end{itemize}
            \item \underline{Step 3}: Submit the resulting evolution tasks described and the architectural description evolved; and
            \item \underline{Step 4}: Answer the qualitative analysis evaluation form. 
        \end{itemize}
    

In particular, to evaluate the ease and intent to use of both guidelines and ad hoc approach (i.e., ``support method'' used by them), statements (questions)  based on TAM (Technology Acceptance Model) \cite {davis1989user} as presented in Table \ref{questoes_tam} were included in the evaluation form. Respondents used Likert scale to score these statements: ``Totally disagree'' (1), ``Partially disagree'' (2) ``Neutral'' (3), ``Partially agree'' (4), and ``Totally agree'' (5).

\begin{table}[!h]\footnotesize
\centering
\caption{Statements included in the evaluation form to evaluate ease and intent to use of guidelines and ad hoc approach}
\label{questoes_tam}
\begin{tabular}{|c|l|}
\hline
\rowcolor[HTML]{C0C0C0} 
\multicolumn{1}{|l|}{\cellcolor[HTML]{C0C0C0}} & \multicolumn{1}{c|}{\cellcolor[HTML]{C0C0C0}\textbf{Ease of Use (EU)}} \\ \hline
EU1                                  & \begin{tabular}[c]{@{}l@{}} Learning the ``support method'' was easy to me.\end{tabular} \\ \hline
EU2                                  & \begin{tabular}[c]{@{}l@{}}The structure of the ``support method'' is \\ clear and  understandable to me.\end{tabular} \\ \hline
EU3                                  & \begin{tabular}[c]{@{}l@{}}It was easy to perform the evolution of the \\ reference architecture description.\end{tabular}\\ \hline
EU4                                  & \begin{tabular}[c]{@{}l@{}}It was easy to understand and use the ``support \\ method''.\end{tabular}\\ \hline
\rowcolor[HTML]{C0C0C0} 
\textbf{}                                      & \multicolumn{1}{c|}{\cellcolor[HTML]{C0C0C0}\textbf{Intention to Use (IU)}}\\ \hline
IU1                                   & \begin{tabular}[c]{@{}l@{}}Considering that I have to evolve reference \\ architectures description with frequency, \\ I intend to use the ``method of support ''. \end{tabular}    \\ \hline
\end{tabular}
\end{table}

To evaluate the efficiency of the ``support method'', we analyzed the architectural descriptions developed by each participant to observe if each evolution request was fulfilled taking into account the completeness and correctness of the description obtained. To record this, we used the following Likert scale: ``Not performed'', ``Performed, but not satisfactory'', ``Partially satisfactory'' or ``Satisfactory''.

\subsection{Execution and Results}


After conducting an full-day training, we performed a full-day qualitative analysis with two participants of \emph{G-Guidelines} absent. Hence, this group had seven participants, while \emph{G-Ad-Hoc} had nine, totaling 16 participants. Executed according to the planning, 
participants performed all steps and results were collected and are synthesized following.



\vspace{.7cm}
\noindent\textbf{Ease and Intention to Use}


Answered by all participants, responses to the evaluation form were summarized by calculating the median value as presented in Table \ref{my-label}. Median was adopted because it is a significant statistical measure applicable to ordinal scales \cite{wohlin2012experimentation}, as we had ordinal ones (i.e., 1 to 5 according to Likert scale).


On average, we observed the guidelines were better evaluated compared to the ad hoc approach in terms of both ease to use and intention to use. It was also noted \emph{G-Guidelines} was neutral regarding the guidelines were easy to learn. We believe this is because our guidelines present a detailed documentation, what is positive considering the need of systematizing evolution tasks, but time-consuming for the first users (as the case of our participants). In addition, this group partially agreed with the guidelines' structure and their use, as well as the ease to perform architectural evolution. Regarding the intention to use, in general, the guidelines were better scored than the ad-hoc approach. On the other hand, \emph{G-Ad-hoc} disagreed totally or partially that the ad-hoc approach was easy to understand and use.
Hence, there is some evidence that the guidelines could ease software architects in the evolution of reference architecture descriptions, better than an ad hoc approach.

\begin{table}[h] \footnotesize
\centering
\caption{Median by Method Used}
\label{my-label}
\begin{tabular}{lcc}
\rowcolor[HTML]{9B9B9B} 
              & \multicolumn{1}{l|}{\cellcolor[HTML]{9B9B9B}\textit{\textbf{Ad hoc}}} & \multicolumn{1}{l}{\cellcolor[HTML]{9B9B9B}\textbf{Guidelines}} \\ \hline
\rowcolor[HTML]{C0C0C0} 
              & \multicolumn{2}{c}{\cellcolor[HTML]{C0C0C0}\textbf{Median}}                                                                            \\ \hline
\rowcolor[HTML]{EFEFEF} 
\multicolumn{3}{c}{\cellcolor[HTML]{EFEFEF}\textbf{Ease to Use (EU)}}                                                                        \\ \hline
EU1 & \multicolumn{1}{c|}{1}                                                & 3                                                               \\
EU2 & \multicolumn{1}{c|}{2}                                                & 4                                                               \\
EU3 & \multicolumn{1}{c|}{1}                                                & 4                                                               \\
EU4 & \multicolumn{1}{c|}{2}                                                & 4                                                               \\ \hline
\rowcolor[HTML]{EFEFEF} 
\multicolumn{3}{c}{\cellcolor[HTML]{EFEFEF}\textbf{Intention to Use (IU)}}                                                                                    \\ \hline
IU1  & \multicolumn{1}{c|}{3}                                                & 4                                                               \\\hline
\end{tabular}
\end{table}

\vspace{0.3cm}
\noindent\textbf{Efficiency} 
 


Table \ref{eficiencia_porcentagem} summarizes how participants performed the three evolution requests (R1, R2, and R3). 
Considering R1, a similar amount of participants of both groups did not perform this request, while \emph{G-Ad-Hoc} satisfactorily or partially satisfactorily performed it compared to \emph{G-Guidelines}. We believe this is because R1 had low complexity what did not require a systematized means as the guidelines are. 
Considering R2 and R3 that had average complexity, most \emph{G-Ad-Hoc} participants were unable to comply them; on the other hand, \emph{G-Guidelines} had better results in both requests; i.e., 
participants satisfactorily or partially satisfactorily performed them (43\% and 29\%). Analyzing the percentage of participants in this group who did not perform R2 and R3 (57\% in both), it can be observed the extensive documentation provided for this group may have contributed to this result. Generally speaking, these results suggest the guidelines may be more efficient for executing more complex requests.

\begin{table}[h]
\centering
\caption{Percentage of participants regarding compliance with evolution requests (\textit{G-Ad-hoc} x \emph{G-Guidelines})}
\footnotesize
\label{eficiencia_porcentagem}
\begin{tabular}{lcccccc}
                                                                              & \multicolumn{3}{c|}{\cellcolor[HTML]{9B9B9B}\textit{\textbf{Ad hoc}}}                                                             & \multicolumn{3}{c}{\cellcolor[HTML]{9B9B9B}\textbf{Guidelines}}                                             \\ \cline{2-7} 
                                                                              & \multicolumn{6}{c}{\cellcolor[HTML]{C0C0C0}\textbf{Evolution Request}}                                                                                                                                                                                      \\ \cline{2-7} 
                                                                              & \cellcolor[HTML]{EFEFEF}\textbf{R1} & \cellcolor[HTML]{EFEFEF}\textbf{R2} & \multicolumn{1}{c|}{\cellcolor[HTML]{EFEFEF}\textbf{R3}} & \cellcolor[HTML]{EFEFEF}\textbf{R1} & \cellcolor[HTML]{EFEFEF}\textbf{R2} & \cellcolor[HTML]{EFEFEF}\textbf{R3} \\ \hline
Not performed                                                         & 44\%                               & 89\%                               & \multicolumn{1}{c|}{89\%}                               & 42\%                               & 57\%                               & 57\%                               \\\hline
Performed, but \\not satisfactory                                                     & 0\%                                & 11\%                               & \multicolumn{1}{c|}{0\%}                                & 0\%                                & 0\%                                & 14\%                               \\\hline
\begin{tabular}[c]{@{}l@{}}Partially \\ satisfactory\end{tabular} & 0\%                                & 0\%                                & \multicolumn{1}{c|}{0\%}                                & 29\%                               & 0\%                                & 0\%                                \\\hline
Satisfactory                                                         & 56\%                               & 0\%                                & \multicolumn{1}{c|}{11\%}                               & 29\%                               & 43\%                               & 29\%                               \\ \hline
\end{tabular}
\end{table}


\section{Discussions}\label{sec:discussion}

By describing in details our guidelines and their rules, they could be applied to systematically and properly evolve descriptions of reference architectures. Besides, these guidelines could be adapted to each evolution scenario (e.g., a specific architecture) in the sense that a subset of them could be selected based on, for instance, elements needed to be addressed; for example, a given architecture requiring a more complete set of architectural views could be updated applying guidelines related to element Architecture View. Having this possibility of adaptation, we also foresee their use in architectures with different description level (from single informal description to complete documentation) or different application domains, even those critical ones. In particular, guidelines related to ``Quality'' element specifically deal with evolution of quality issues. Being independent from architectural description languages (ADL), our guidelines can also address architectures represented in different ADL or a set of them.


Based on a well-known international standard for architecture description (in the case, ISO/IEC/IEEE 42010) and also founded in a large and consolidated literature, our guidelines can be considered correct by construction. Specifically, these guidelines were based on 
a reference architecture evaluation checklist (in the case, FERA); however, since this checklist adopted ISO/IEC/IEEE 42010 as its main foundation, we believe 
our guidelines cover all main evolution points in descriptions of reference architectures.  Additionally, we conducted a qualitative analysis to 
illustrate how these guidelines could work compared with an ad hoc way. %
As some participants in our analysis (six of them) worked in industry, i.e., they were also practitioners, we believe our findings are relevant to practitioners who work with reference architectures and are concerned with systematically evolving the description of their architectures.
In addition, according to Salman et al. \cite{Salman2015}, the outcomes found by students in software engineering experiments are similar to those found by practitioners when both use/apply a novel approach, method or technique being proposed, as it is the case of guidelines proposed in this paper and used in the qualitative analysis. Although results cannot be generalized, in general, our guidelines seem to be easier to be used and are more efficient than an ad hoc one; furthermore, a higher intention to adopt them was similarly observed. 

\subsection{Threats to Validity}
The main threats to validity of the qualitative analysis are described below:

\begin{itemize}
\setlength\itemsep{-0.05em}

    \item \textbf{Internal Validity}: Regarding this validity, we observed: (i) Different level of experience among participants (to mitigate this threat, training on software architecture and reference architecture was provided for all participants; besides, two well-balanced groups were formed based on profile form); (ii) Low level of performance of participants (to mitigate it, the training and analysis execution were conducted in different days); and (iv) Structure of the documents provided (to mitigate it, before executing analysis, we conducted a pilot study to identify problems and improve all documents).

    \item \textbf{Construct Validity}: Cambuci documentation used in our work could lead to different interpretations; therefore, to mitigate this threat, a detailed explanation and discussion sessions were provided to all participants during training day. 
    
    \item \textbf{External Validity}: With regard to this validity: (i) Results obtained cannot be generalized, due to small number of participants; 
    (ii)  Participants may not be representative of the general population: to mitigate it, we only selected graduate students in Computer Science with some experience in software design, including six of them who were also working in industry and had experience in software system design. Besides that, training was 
    provided to reduce the difference of knowledge among 
    them; and (iii) Evolution tasks of reference architecture descriptions may not be representative of real-world cases: to mitigate it, we considered three tasks with distinct complexity levels and with situations that could occur in a real-world case.
    
    \item \textbf{Conclusion Validity}: The number of participants did not make possible applying statistical tests and also did not allow us to generalize results. In this way, it was only possible to observe even without statistical significance a trend to the efficiency and ease to use our guidelines.
\end{itemize}

\subsection{Perspectives of Future Work}

Considering the novelty of this work, it opens a number of perspective for future work. First of all, a supporting tool is essential to assist search and retrieval of most appropriate guidelines in each case and also to disseminate them. Another essential step is to conduct empirical studies with a larger number of participants from both academia and industry to accurately evaluate efficiency, ease, and intention to use the guidelines to confirm results, including with statistical significance. Whether conducted on known reference architectures, as Autosar, such studies could bring evidences on  real impacts of our guidelines also in industry architectures. 

Most existing reference architectures are still described in an informal way (boxes and lines, for instance) and in some cases in UML together with textual descriptions. Sections ``how to represent'' of our guidelines and sections ``actions'' of evolution rules could be refined to guide choice of ADL and their models, even formal ones, what could enable formal verification by specific tools. 


There is also a couple of approaches to systematically design reference architectures such as ProSA-RA, but no one explicitly incorporates evolution issues. Hence, another investigation strand is to incorporate our guidelines in such approaches aiming at becoming them more complete solutions, followed by empirical studies to demonstrate their effectiveness.

\section{Conclusion}
\label{sec:conclusion}

With the number of software systems continuously increasing, reference architectures have gained their importance, including in industry scenario developing critical, complex systems. Much effort has been already devoted to advance both the state of the practice and state of the art with theoretical foundation and automated solutions to the reference architecture field, also there are an expressive number of more than 150\footnote{We found 161 reference architectures through a systematic mapping study conducted in February, 2018.} reference architectures in different application domains already proposed. Unexpectedly, most of these architectures have not been updated and there are also few studies investigating the evolution of reference architectures and their description. 

This work brought a way to update description of reference architectures. Correct by construction by considering large and consolidated literature and practice, together with our experience for years researching and designing reference architectures, this work particularly present a set of systematic steps organized in the format of guidelines. We believe our correct-by-construction guidelines are a novelty step towards contributing to meet one of the main factors to sustain reference architectures, i.e., their adequate description \cite{Volpato2017b} and, as a result, achieve what was previously referred as sustainable reference architectures~\cite{Venters18SSRP}.

With this work, we intend to call attention and change the mindset of the community to this important task of evolving reference architectures and their descriptions, aiming to keep these architectures useful and sustainable along the time, fulfilling their main role of reusing knowledge.

\section*{Acknowledgement} 
This work was supported by Federal University of Mato Grosso do Sul and Brazilian funding agencies CAPES (Finance Code 001); FAPESP (grants: 2015/24144-7); and CNPq (grants: 313245/2021-5).

\bibliography{references}

\newpage
\setcounter{table}{0}
\renewcommand{\thetable}{A.\arabic{table}}
\appendix

\section{Examples of Guidelines}\label{other_guidelines}

This appendix presents five other guidelines to evolve reference architecture descriptions. For the sake of space, the part \textbf{``Evolution tasks and rules''} of each guideline is summarized.

\begin{table*}[h]
\footnotesize 
	\centering
	\caption{Guideline D\_Acquisition\_1}\
	\label{dAcquisition1}
	\begin{tabular}{|l|l|}
		\hline
		\rowcolor[HTML]{9B9B9B} 
		\textbf{Guideline}                                                                                     & \begin{tabular}[c]{@{}l@{}} D\_Acquisition\_1\end{tabular}                                                                                                             \\ \hline
		\rowcolor[HTML]{C0C0C0} 
		\textbf{Element}                                                                                     & Acquisition                                              \\ \hline
		\textbf{\begin{tabular}[c]{@{}r@{}}Applicable to \\FERA question \end{tabular}}                  & \begin{tabular}[c]{p{0.8\textwidth}}2-25; 3-51
        \end{tabular}                                                                \\ \hline
		\textbf{What to do}                                                                                  & \begin{tabular}[c]{p{0.8\textwidth}}
	        To present or detail information regarding the acquisition process to which the reference architecture is involved.
		\end{tabular}        
		\\ \hline
		\textbf{\begin{tabular}[c]{@{}r@{}}How to represent\end{tabular}}  & 
		
		\begin{tabular}[c] {p{0.8\textwidth}}
		\textbf{Textual}: To present a section in the general description of the reference architecture namely ``Acquisition'', related to the acquisition process, to detail how the reference architecture may be acquired. For this, we suggest consulting the Acquisition Process of ISO/IEC 12208 \cite{iso12207:2008} and CMMI-ACQ \cite{SEI2010}.
		\end{tabular}                           \\\hline
		\textbf{How to make}                                                                                   & \begin{tabular}[c]{p{0.8\textwidth}}
		Execute step ``Architectural Analysis'' when there is no information source related to acquisition of software product. \\
		Execute step ``Architectural Synthesis'' based on evolution tasks of interest: Addition or Modification.
		\end{tabular} \\ \hline
	
		& \begin{tabular}[c]{p{0.8\textwidth}}
		Addition: R-aq-1 (To create a section named ``Acquisition Process'' in the general description of the reference architecture and describe in detail how the reference architecture acquisition process will be done, monitored and evaluated)
\end{tabular}      \\ \cline{2-2} 
		
		& \begin{tabular}[c]{p{0.7\textwidth}}
		Removal: --
		\end{tabular}       \\ \cline{2-2} 
		\multirow{-3}{*}{\textbf{\begin{tabular}[c]{@{}r@{}} Evolution tasks \\
		and rules\end{tabular}}} & 
		\begin{tabular}[c]{p{0.8\textwidth}}
		Modification: R-aq-2  (To perform the necessary changes in the process acquisition description in  Section ``Acquisition Process'' to suitably portray its realization, monitoring and evaluation)
		\end{tabular} \\ \hline
		\textbf{\begin{tabular}[c]{@{}r@{}}Artifact  involved \\  \end{tabular}}                              & \begin{tabular}[c]{p{0.8\textwidth}}Architectural Description \end{tabular}  \\ \hline
	\end{tabular}
\end{table*}

\begin{table*}[h]
\footnotesize 
	\centering
	\caption{Guideline D\_Quality\_4}\
	\label{dQuality4}
	\begin{tabular}{|l|l|}
		\hline
		\rowcolor[HTML]{9B9B9B} 
		\textbf{Guideline}                                                                                     & \begin{tabular}[c]{@{}l@{}} D\_Quality\_4\end{tabular}                                                                                                             \\ \hline
		\rowcolor[HTML]{C0C0C0} 
		\textbf{Element}                                                                                     & Quality                                              \\ \hline
		\textbf{\begin{tabular}[c]{@{}r@{}}Applicable to \\FERA question \end{tabular}}                  & \begin{tabular}[c]{p{0.8\textwidth}}2-25; 3-17
        \end{tabular}                                                                \\ \hline
		\textbf{What to do}                                                                                  & \begin{tabular}[c]{p{0.8\textwidth}}
	        To present the characteristics and sub-characteristics of the quality attributes  mapped to each architectural requirement of the reference architecture.
		\end{tabular}        
		\\ \hline
		\textbf{\begin{tabular}[c]{@{}r@{}}How to represent\end{tabular}}  & 
		
		\begin{tabular}[c] {p{0.8\textwidth}}
		\textbf{Textual}: To describe the characteristics and sub-characteristics of the quality attributes mapped for each architectural requirement of the reference architecture (RA) in Section ``Architectural Description'' of Section ``Domain Data'' of the reference architecture description, as exemplified below.
		\\
		\\
		\begin{tabular}{|
>{\columncolor[HTML]{FFFFFF}}l |
>{\columncolor[HTML]{FFFFFF}}l |
>{\columncolor[HTML]{FFFFFF}}l |
>{\columncolor[HTML]{FFFFFF}}l |}
\hline
\multicolumn{1}{|c|}{\cellcolor[HTML]{FFFFFF}{\color[HTML]{222222} \textbf{ID}}} & \multicolumn{1}{c|}{\cellcolor[HTML]{FFFFFF}{\color[HTML]{222222} \textbf{Architectural Requirement}}} & \multicolumn{1}{c|}{\cellcolor[HTML]{FFFFFF}{\color[HTML]{222222} \textbf{Concept}}} & \multicolumn{1}{c|}{\cellcolor[HTML]{FFFFFF}{\color[HTML]{222222} \textbf{\begin{tabular}[c]{@{}c@{}}Quality Characteristic/\\ Sub-characteristic\end{tabular}}}} \\ \hline
{\color[HTML]{222222} RA1}                                                       & {\color[HTML]{222222} The RA should allow  ...}                                                        & {\color[HTML]{222222} ConceptA}                                                      & {\color[HTML]{222222} Maintainability/Modularization}                                                                                                             \\ \hline
{\color[HTML]{222222} RA2}                                                       & {\color[HTML]{222222} ...}                                                                             & {\color[HTML]{222222} ...}                                                           & {\color[HTML]{222222} ...}                                                                                                                                        \\ \hline
\end{tabular}
\\
\\
		\end{tabular}                           \\\hline
		\textbf{How to make}                                                                                   & \begin{tabular}[c]{p{0.8\textwidth}}
		Execute step ``Architectural Analysis'' when there is no information source related to characteristics and sub-characteristics of the quality attributes of the reference architecture.  \\
		Execute step ``Architectural Synthesis'' based on evolution tasks of interest: Modification.
		\end{tabular} \\ \hline
	
		& \begin{tabular}[c]{p{0.8\textwidth}}
		Addition: --
\end{tabular}      \\ \cline{2-2} 
		
		& \begin{tabular}[c]{p{0.7\textwidth}}
		Removal: --
		\end{tabular}       \\ \cline{2-2} 
		\multirow{-3}{*}{\textbf{\begin{tabular}[c]{@{}r@{}} Evolution tasks \\
		and rules\end{tabular}}} & 
		\begin{tabular}[c]{p{0.8\textwidth}}
		Modification: R-dd-3  (To perform the necessary changes in the description of  characteristics and sub-characteristics of the quality attributes  mapped to each architectural requirement of the reference architecture in  Section ``Architectural Description`` of Section ``Domain Data'' of the reference architecture description to suitably portray necessary adjustments or internal and/or external demands)
		\end{tabular} \\ \hline
		\textbf{\begin{tabular}[c]{@{}r@{}}Artifact  involved \\  \end{tabular}}                              & \begin{tabular}[c]{p{0.8\textwidth}}Architectural Description \end{tabular}  \\ \hline
	\end{tabular}
\end{table*}

\begin{table*}[h]
\footnotesize 
	\centering
	\caption{Guideline D\_Technical\_Solution\_2}\
	\label{dTechnicalSolution2}
	\begin{tabular}{|l|l|}
		\hline
		\rowcolor[HTML]{9B9B9B} 
		\textbf{Guideline}                                                                                     & \begin{tabular}[c]{@{}l@{}} D\_Technical\_Solution\_2\end{tabular}                                                                                                             \\ \hline
		\rowcolor[HTML]{C0C0C0} 
		\textbf{Element}                                                                                     & Technical Solution                                              \\ \hline
		\textbf{\begin{tabular}[c]{@{}r@{}}Applicable to \\FERA question \end{tabular}}                  & \begin{tabular}[c]{p{0.8\textwidth}}2-25; 3-28
        \end{tabular}                                                                \\ \hline
		\textbf{What to do}                                                                                  & \begin{tabular}[c]{p{0.8\textwidth}}
	        To discriminate OTS (Off-The-Shelf) and/or OSS (Open Source Software) components needed  to implement parts of the reference architecture, if it is possible to  use them.
		\end{tabular}        
		\\ \hline
		\textbf{\begin{tabular}[c]{@{}r@{}}How to represent\end{tabular}}  & 
		
		\begin{tabular}[c] {p{0.8\textwidth}}
		\textbf{Textual}: To discriminate each part of the reference architecture that can be implemented using  OTS (Off-The-Shelf) and OSS (Open Source Software) components in Section ``Technical Solution'' (several parts can be implemented by only one component) of the reference architecture description, as well as the provider and the dependency among the components (if there is), as exemplified below.
	        \\
	        \\
	        \begin{tabular}{|
>{\columncolor[HTML]{FFFFFF}}l |
>{\columncolor[HTML]{FFFFFF}}l |
>{\columncolor[HTML]{FFFFFF}}l |
>{\columncolor[HTML]{FFFFFF}}l |}
\hline
\multicolumn{1}{|c|}{\cellcolor[HTML]{FFFFFF}{\color[HTML]{222222} \textbf{Part of RA}}} & \multicolumn{1}{c|}{\cellcolor[HTML]{FFFFFF}{\color[HTML]{222222} \textbf{OTS and/or OSS components used}}} & \multicolumn{1}{c|}{\cellcolor[HTML]{FFFFFF}{\color[HTML]{222222} \textbf{Provider}}} & \multicolumn{1}{c|}{\cellcolor[HTML]{FFFFFF}{\color[HTML]{222222} \textbf{Dependency among components}}} \\ \hline
{\color[HTML]{222222} Part X}                                                            & {\color[HTML]{222222} Component A}                                                                         & {\color[HTML]{222222} Provider 1}                                                     & {\color[HTML]{222222} -}                                                                                 \\ \hline
{\color[HTML]{222222} Part Y}                                                            & {\color[HTML]{222222} Component B, Component C}                                                            & {\color[HTML]{222222} Provider 2}                                                     & {\color[HTML]{222222} Component C depends on Component B}                                                \\ \hline
\end{tabular}
	        \\
	        \\
		
		\end{tabular}                           \\\hline
		\textbf{How to make}                                                                                   & \begin{tabular}[c]{p{0.8\textwidth}}
		Execute step ``Architectural Analysis'' when there is no information source related to possible components that could be used to implement parts of the reference architecture. 
		 \\
		Execute step ``Architectural Synthesis'' based on evolution tasks of interest: Addition or Modification.
		\end{tabular} \\ \hline
	
		& \begin{tabular}[c]{p{0.8\textwidth}}
		Addition: R-st-5 (To discriminate the components used to implement the reference architecture in Section ``Technical Solution'' of the reference architecture description, as presented in ``How to represent'')
\end{tabular}      \\ \cline{2-2} 
		
		& \begin{tabular}[c]{p{0.7\textwidth}}
		Removal: --
		\end{tabular}       \\ \cline{2-2} 
		\multirow{-3}{*}{\textbf{\begin{tabular}[c]{@{}r@{}} Evolution tasks \\
		and rules\end{tabular}}} & 
		\begin{tabular}[c]{p{0.8\textwidth}}
		Modification: R-st-6  (To perform the necessary changes in the discrimination of  components used to implement the reference architecture in Section ``Technical Solution'' of the reference architecture description  to suitably portray necessary adjustments or demands)
		\end{tabular} \\ \hline
		\textbf{\begin{tabular}[c]{@{}r@{}}Artifact  involved \\  \end{tabular}}                              & \begin{tabular}[c]{p{0.8\textwidth}}Architectural Description \end{tabular}  \\ \hline
	\end{tabular}
\end{table*}

\begin{table*}[h]
\footnotesize 
	\centering
	\caption{Guideline D\_Domain\_Data\_4}\
	\label{dDomainData4}
	\begin{tabular}{|l|l|}
		\hline
		\rowcolor[HTML]{9B9B9B} 
		\textbf{Guideline}                                                                                     & \begin{tabular}[c]{@{}l@{}} D\_Domain\_Data\_4\end{tabular}                                                                                                             \\ \hline
		\rowcolor[HTML]{C0C0C0} 
		\textbf{Element}                                                                                     & Domain Data                                              \\ \hline
		\textbf{\begin{tabular}[c]{@{}r@{}}Applicable to \\FERA question \end{tabular}}                  & \begin{tabular}[c]{p{0.8\textwidth}}2-25; 3-12
        \end{tabular}                                                                \\ \hline
		\textbf{What to do}                                                                                  & \begin{tabular}[c]{p{0.8\textwidth}}
	        To clearly describe the objective, scope, target domain of the reference architecture, as well as the context in which the reference architecture can be instantiated.
		\end{tabular}        
		\\ \hline
		\textbf{\begin{tabular}[c]{@{}r@{}}How to represent\end{tabular}}  & 
		
		\begin{tabular}[c] {p{0.8\textwidth}}
		\textbf{Textual}:To clearly describe  the objective, scope,  target domain and context in which the reference architecture can be instantiated in Section ``Introduction'' of the reference architecture description. \\
	    \textbf{Graphical}: If necessary, to draw a Venn diagram to show the limit of the reference architecture target domain regarding the neighborhood domains. 
		\end{tabular}                           \\\hline
		\textbf{How to make}                                                                                   & \begin{tabular}[c]{p{0.8\textwidth}}
		Execute step ``Architectural Analysis'' when there is no information source related to objective, scope, target domain and instantiation context of the reference architecture.  
		 \\
		Execute step ``Architectural Synthesis'' based on evolution tasks of interest: Addition or Modification.
		\end{tabular} \\ \hline
	
		& \begin{tabular}[c]{p{0.8\textwidth}}
		Addition: R-da-4 (To clearly describe the objective, scope,  target domain and context in which the reference architecture can be instantiated in Section ``Introduction'' of the reference architecture description) 
\end{tabular}      \\ \cline{2-2} 
		
		& \begin{tabular}[c]{p{0.7\textwidth}}
		Removal: --
		\end{tabular}       \\ \cline{2-2} 
		\multirow{-3}{*}{\textbf{\begin{tabular}[c]{@{}r@{}} Evolution tasks \\
		and rules\end{tabular}}} & 
		\begin{tabular}[c]{p{0.8\textwidth}}
		Modification: R-da-5 (To perform the necessary changes in the description the objective, scope,  target domain and context in which the reference architecture can be instantiated in Section ``Introduction'' to suitably portray necessary adjustments or internal and/or external demands)
		\end{tabular} \\ \hline
		\textbf{\begin{tabular}[c]{@{}r@{}}Artifact  involved \\  \end{tabular}}                              & \begin{tabular}[c]{p{0.8\textwidth}}Architectural Description \end{tabular}  \\ \hline
	\end{tabular}
\end{table*}

\begin{table*}[h]
\footnotesize 
	\centering
	\caption{Guideline D\_Instantiation\_2}\
	\label{dInstantiation2}
	\begin{tabular}{|l|l|}
		\hline
		\rowcolor[HTML]{9B9B9B} 
		\textbf{Guideline}                                                                                     & \begin{tabular}[c]{@{}l@{}} D\_Instantiation\_2\end{tabular}                                                                                                             \\ \hline
		\rowcolor[HTML]{C0C0C0} 
		\textbf{Element}                                                                                     & Instantiation                                              \\ \hline
		\textbf{\begin{tabular}[c]{@{}r@{}}Applicable to \\FERA question \end{tabular}}                  & \begin{tabular}[c]{p{0.8\textwidth}} 3-13; 3-24
        \end{tabular}                                                                \\ \hline
		\textbf{What to do}                                                                                  & \begin{tabular}[c]{p{0.8\textwidth}}
	        To present the instantiation guidelines for the reference architecture, and/or case studies and examples, by describing an instantiation based on the reference architecture description.
		\end{tabular}        
		\\ \hline
		\textbf{\begin{tabular}[c]{@{}r@{}}How to represent\end{tabular}}  & 
		
		\begin{tabular}[c] {p{0.8\textwidth}}
		\textbf{Textual}: To describe guidelines to help software architects in instantiating reference architectures, in Section ``Instantiation Guidelines'' of Section ``Instantiation'' of the reference architecture description. To include case studies and examples of the reference architecture instantiation in an Appendix.
		\end{tabular}                           \\\hline
		\textbf{How to make}                                                                                   & \begin{tabular}[c]{p{0.8\textwidth}}
		Execute step ``Architectural Synthesis'' based on evolution tasks of interest: Addition or Modification.
		\end{tabular} \\ \hline
	
		& \begin{tabular}[c]{p{0.8\textwidth}}
		Addition: R-ins-1; R-ins-2; R-ins-4 (To describe in detail the guidelines to provide step-by-step instructions on how to instantiate the reference architecture  in Section ``Instantiation Guidelines'' of Section ``Instantiation'' of the reference architecture description. In addition, by including an Appendix containing case studies and examples of the reference architecture instantiation) 
\end{tabular}      \\ \cline{2-2} 
		
		& \begin{tabular}[c]{p{0.7\textwidth}}
		Removal: --
		\end{tabular}       \\ \cline{2-2} 
		\multirow{-3}{*}{\textbf{\begin{tabular}[c]{@{}r@{}} Evolution tasks \\
		and rules\end{tabular}}} & 
		\begin{tabular}[c]{p{0.8\textwidth}}
		Modification: R-ins-3; R-ins-5 (To perform the necessary changes in the instantiation guidelines,  in Section “Instantiation Guidelines” of Section “Instantiation” of the reference architecture description, as well as in the case studies and examples of the reference architecture instantiation, in Appendix, to suitably portray necessary adjustments into the reference architecture itself or to improve the guidelines description to make it more intelligible)
		\end{tabular} \\ \hline
		\textbf{\begin{tabular}[c]{@{}r@{}}Artifact  involved \\  \end{tabular}}                              & \begin{tabular}[c]{p{0.8\textwidth}}Architectural Description \end{tabular}  \\ \hline
	\end{tabular}
\end{table*}

\end{document}